\documentclass{ws-p8-50x6-00}

\begin{document}

\title{Recent Developments in Cosmology}

\author{Andrew R.~Liddle}

\address{Astronomy Centre, University of Sussex, Brighton BN1 9QJ, United 
Kingdom\\ 
E-mail:a.liddle@sussex.ac.uk}

\maketitle

\abstracts{This article gives an overview, aimed at theoretical particle 
physicists, of some recent developments in cosmology.}

\section{Overview}

The last few years in cosmology have been thrilling ones, as dramatic 
improvements in observational technology have begun to impose stringent 
constraints on theoretical ideas in cosmology built up over the preceding two 
decades. For the purposes of this article I'll focus on the following set of 
overall goals:
\begin{itemize}
\item To obtain a physical description of the Universe, including its global 
dynamics and matter content.
\item To measure the cosmological parameters describing the Universe, and to 
develop a fundamental understanding of as many of those parameters as possible.
\item To understand the origin and evolution of cosmic structures.
\item To understand the physical processes which took place during the extreme 
heat and density of the early Universe.
\end{itemize}
Over recent years, much progress has been made on all of these topics, to the 
extent that it is widely believed amongst cosmologists that we may stand on the 
threshold of the first precision cosmology, in which the parameters necessary to 
describe our Universe have been identified and will soon be, in most cases at 
least, measured 
to a satisfying degree of precision. Whether this optimism has any grounding in 
reality remains to be seen, though so far the signs are promising in that the 
basic picture of cosmology, centred around the Hot Big Bang, has time and again 
proven the best framework for interpreting the constantly improving 
observational situation.

In particular, the process of cosmological parameter estimation is well 
underway, thanks to observations of distant Type Ia supernovae, of galaxy 
clustering, and of
the cosmic microwave background. These have established a standard cosmological 
model, where the Universe is dominated by dark energy, contains substantial dark 
matter, and with the baryons from which we are made comprising only around 4\%. 
Overall this model can be described by around ten parameters (e.g.~see 
Ref.~\cite{WTZ}), and the viable region of parameter space is starting to shrink 
under pressure from observations. 
However, it is worth bearing in mind that we seek high precision determinations 
at least in part because they ought to shed light on fundamental physics, and 
there progress has been less rapid. Some parameters are likely to have no 
particular fundamental importance (for instance, there would probably be little 
fundamental significance were the Hubble constant to turn out to be $63 \, {\rm 
km \, s}^{-1} \, {\rm Mpc}^{-1}$ rather than say $72 \, {\rm km \, s}^{-1} \, 
{\rm Mpc}^{-1}$, though accurate determination of this parameter is essential if 
we are to pin down other parameters), but the 10\% or so measured accuracy of 
the baryon density is 
to be set against the lack of even an order-of-magnitude theoretical 
understanding thus far.

This article does not attempt to cover the complete range of moden cosmology, 
but is intended as a status report on a subset of topics which I've chosen as 
being potentially of the most interest to theoretical particle physicists. The 
main descriptive sections concern structure formation in the Universe and the 
inflationary cosmology, and the final section is a mixed bag of especially 
topical subjects.

\section{Structure Formation in the Universe}

\subsection{Gravitational instability}

One of the most powerful tools in cosmology is the development of structures. By 
`structure' I mean anything corresponding to inhomogeneity within the Universe, 
be it galaxies, variations in the gravitational potential, or anisotropies in 
the cosmic microwave background. The evolution of structures proves sensitive to 
all the main cosmological parameters, and hence is well suited to constraining 
them. Different types of observation naturally probe different physical regimes, 
for instance small verses large scales, and also different stages of the 
Universe's evolution, with the microwave background probing the Universe when it 
was around one thousandth of its present size.

The young universe was much closer to uniformity than the present state; for 
instance the irregularities in the cosmic microwave background are only around 
one part in $10^5$, while the present matter distribution features highly 
overdense galaxies with voids in between. The main driving force in this 
evolution, at least in its initial stages, is simply gravity; any initial 
overdensity will exert an unbalanced gravitational force upon neighbouring 
material and will tend to accrete material, amplifying the original 
perturbations. At least until well after the cosmic microwave background 
radiation is released, the perturbation evolution is well described on all 
scales by linear 
perturbation theory, though ultimately linear theory for the density field 
breaks down on short scales 
as virialized galaxies begin to form. On sufficiently large scales linear theory 
remains adequate even today.

The Hot Big Bang model, supplemented by gravitational instability in order to 
form structures, gives an excellent broad-brush description of our Universe. 
However, like any theory or model in physics, its predictions depend on some 
input parameters not specified by the theory. A key goal is to measure those 
parameters to a satisfying degree of accuracy. For example, the detailed process 
of gravitational instability depends on
\begin{itemize}
\item The expansion rate of the Universe (the Hubble parameter).
\item The density of the material providing the gravitational attraction.
\item The physical properties of the material; for example does it only 
experience gravitational attraction, or are other interactions important?
\item The form of the initial perturbations that get the whole structure 
formation process going.
\end{itemize}
Current ideas in cosmology suggest that around 10 parameters may be sufficient 
to describe our Universe. At present, however, we don't even know the complete 
set of important parameters, far less have accurate values for them all. The 
hope is that over the next few years we will both identify the important 
parameters and measure them to high accuracy, in many cases at the percent 
level.

\subsection{Quantifying microwave background anisotropies}

Although the strongest tests of cosmological models will always come from the 
combination of all available data, cosmic microwave background (CMB) 
anisotropies have received much attention lately (and are likely to be the 
single most important tool for constraining inflation, as discussed in the next 
section), and so it is worth spending some time 
defining the necessary terminology.

We observe the temperature $T(\theta,\phi)$ coming from different directions. We 
write this as a dimensionless perturbation and expand in spherical harmonics
\begin{equation}
\frac{T(\theta,\phi) - \bar{T}}{\bar{T}} = \sum_{\ell,m} a_{\ell m} \, 
Y^{\ell}_m(\theta,\phi) \,.
\end{equation}
There is no unique prediction for the coefficients $a_{\ell m}$, but in the 
simplest inflationary cosmologies they 
are drawn from a gaussian distribution whose mean square is independent of $m$ 
and given by the {\bf radiation angular power spectrum}
\begin{equation}
C_\ell = \left\langle \left|a_{\ell m} \right|^2 \right\rangle_{{\rm ensemble}}
\end{equation}
The ensemble average represents the theorist's ability to average over all 
possible observers in the Universe (or indeed over different quantum mechanical 
realizations), whereas an observer's highest ambition is to estimate it by 
averaging over the multipoles of different $m$ as seen at our own location.
The radiation angular power spectrum depends on all the cosmological parameters, 
and so it can be used to constrain them. To extract the full information 
polarization also has to be measured; this gives three additional power spectra, 
describing two independent modes of polarization, and the cross-correlation 
between the temperature anisotropies and one polarization mode (other 
cross-correlations vanish assuming absence of parity violation).

\begin{figure}[t]
\centering 
\leavevmode\epsfysize=8cm \epsfbox{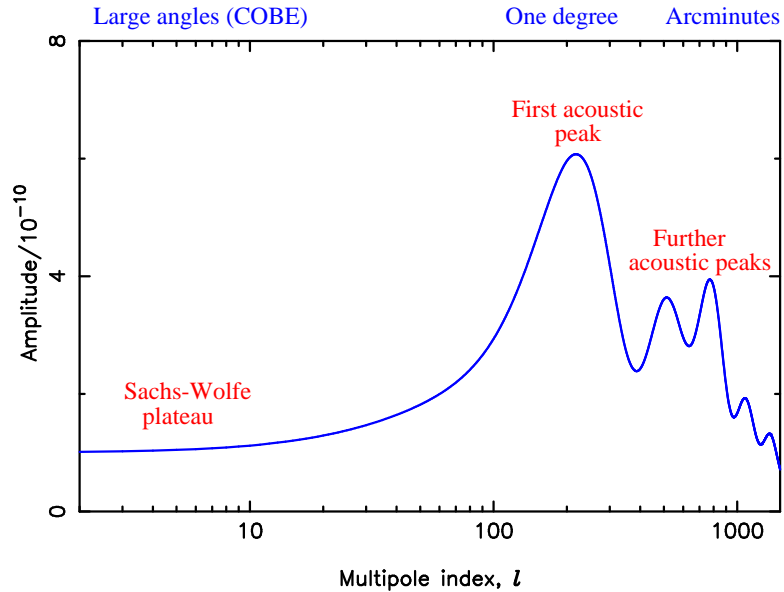}\\ 
\caption[scdm]{\label{f:scdm} The radiation angular power spectrum for a 
particular cosmological model. The annotations name the different features.}
\end{figure}

Computation of the power spectra requires a lot of physics: gravitational 
collapse, photon--electron interactions (and their polarization dependence), 
neutrino free-streaming etc. But as long as the perturbations are small, linear 
perturbation theory can be used which makes accurate calculations possible. A 
major 
step forward for the field was the public release of Seljak \& Zaldarriaga's 
code {\sc cmbfast} \cite{SZ} which can compute the spectra within one percent 
accuracy for a given 
cosmological model 
in around one minute. An example spectrum is shown in Figure~\ref{f:scdm}.

\subsection{Recent CMB results}

During 2000 and 2001 studies of microwave anisotropies took a huge leap forward 
with the first results from a new generation of instruments. First out with 
results was the Boomerang collaboration \cite{Boom}, followed closely by the 
Maxima collaboration
\cite{Max}; these made the first accurate mapping of the first peak in the 
angular power spectrum, corresponding to the first gravitational compression of 
the primordial fluid. The location of this peak is fixed primarily by the 
propagation of light to us after last-scattering, and is a sensitive probe of 
the geometry of the Universe. These results are consistent with a flat geometry, 
with only a small margin for error, and provided a convincing exclusion of a 
low-density open Universe with $\Omega_0 \sim 0.3$ which had up until then been 
regarded as a viable cosmology.

\begin{figure}[t]
\centering 
\leavevmode\epsfysize=8cm \epsfbox{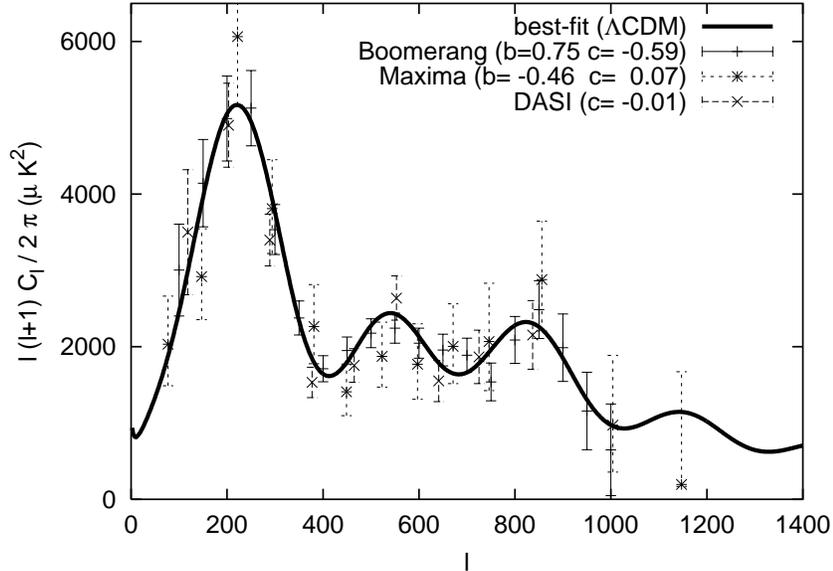}\\ 
\caption[datafig]{\label{f:datafig} A best-fit cosmology with a cosmological 
constant is shown in comparison to recent CMB anisotropy results. The data sets 
of Boomerang, Maxima and DASI are shown with the best-fit calibration values for 
those experiments presuming this best-fit model is correct. [Figure courtesy of 
Julien Lesgourgues.]}
\end{figure}

The first Boomerang and Maxima results gave tentative, but inconclusive, 
indication of further features to small angular scales. The situation improved 
further in mid 2001, with new results from the DASI experiment \cite{DASI} and a 
reanalysis of the Boomerang data \cite{Boom01} including a much larger fraction 
of the total dataset. These results are shown in Figure~\ref{f:datafig}, 
alongside a best-fitting theoretical model. These latest results show the first 
clear evidence for further oscillations in the angular power spectrum, as 
predicted in Figure~\ref{f:scdm}. This observation is of particular qualitative 
significance for the inflationary cosmology, as discussed in the next section, 
and of quantitative significance for constraining the baryon density as 
described in the following subsection.

\newpage
\subsection{The Standard Cosmological Model}

The observations of the last few years have led to the establishment of a 
standard cosmological model, with ingredients as follows.
\begin{quote}
{\large
\begin{tabbing}
\hspace*{0.5cm} \= Cosmological constant \hspace*{1cm} \= $\sim$ 66\%\\
\>Cold dark matter \>$\sim$ 30\%\\
\>Baryons \>$\sim$ 4\%\\
\>Photons and neutrinos \>$\sim 10^{-4}$\\
\>Spatial flatness.\\
\>Hubble constant around $70 \, {\rm km} \, {\rm s}^{-1} \, {\rm Mpc}^{-1}$.\\
\>Initial conditions seeded by slow-roll inflation.
\end{tabbing}
}
\end{quote}
\noindent
This model is in remarkable agreement with observational data.

\subsection*{Baryons}

There are now three independent powerful ways of estimating the baryon density 
of the Universe. Listing the uncertainties at 95\% confidence, we have
\begin{description}
\item{Nucleosynthesis:}~$\Omega_{{\rm baryon}} h^2 = 0.019 \pm 0.002$.\\
It is widely, though not universally, thought that the measurement of the 
deuterium abundance in high-redshift absorption systems gives a highly-accurate 
probe of the baryon density during nucleosynthesis.
\item{Microwave background:}~$\Omega_{{\rm baryon}} h^2 = 0.02 \pm 0.01$.\\
The baryon density can be inferred from the CMB spectrum, as it governs the 
relative heights of the first and second peaks (corresponding to compressions 
and rarefactions of the cosmic fluid respectively). While the Boomerang 2000 
results gave a suspiciously high value for this, reanalysis in 2001 plus new 
results from DASI have brought the value into excellent agreement with 
nucleosynthesis.
\item{Cluster baryon fraction:}~$\Omega_{{\rm baryon}}/(\Omega_{{\rm 
cdm}}+\Omega_{{\rm baryon}}) 
= 0.12 \pm 0.05$.\\
Clusters are observed via X-ray emission from hot intracluster gas. This gas is 
in hydrostatic 
equilibrium against gravity which is principally supplied by the dark matter. 
For the standard cosmological model this agrees excellently with 
nucleosynthesis.
\end{description}

\subsection*{Cosmological constant}

Famously, in 1998 two teams studying distant supernovae discovered that they 
were fainter than expected, and having eliminated other possible causes 
concluded that this was due to the expansion history of the Universe, and 
required a presently-accelerating cosmology \cite{Sn}. This can be brought about 
by a 
cosmological constant $\Lambda$, and if one additionally restricts to a flat 
geometry 
as motivated by the CMB, this leads to the cosmological constant density of the 
Standard Cosmological Model.

Now, if that was the sole evidence for a cosmological constant I wouldn't 
believe it for a second. However the circumstantial evidence is extremely 
powerful; for instance
\begin{enumerate}
\item Microwave anisotropies show the Universe is flat (or close to flat), 
provided the initial perturbations are adiabatic.
\item Nucleosynthesis plus the cluster baryon fraction imply $\Omega_{{\rm 
cdm}}+\Omega_{{\rm baryon}} \sim 0.3$ which implies $\Omega_\Lambda \sim 0.7$.
\item The correct galaxy power spectrum is reproduced if $(\Omega_{{\rm 
cdm}}+\Omega_{{\rm baryon}}) 
h \simeq 0.2$ (where $h$ is the Hubble constant in the usual units); this 
concurs well with direct measures of $h$.
\end{enumerate}
As a result of this and other arguments, the so-called $\Lambda$CDM model 
presently has no serious rivals.

The cosmological constant poses the twin problems of its unexpectedly small 
magnitude (in fundamental physics terms) and the mystery of why it should only 
come to dominate the Universe at the present epoch (around redshift 0.3). To 
address these, instead of a pure cosmological constant, one might prefer an 
effective one, for example a slowly-rolling  potential-dominated scalar field as 
described in the next section for early Universe inflation. Such scenarios are 
known as quintessence. Current observations force such scenarios to be quite 
close to the pure cosmological constant, and though differences may yet be 
unveiled by improved experiments it appears only quite limited information will 
be available. It is actually becoming quite hard to construct simple 
quintessence models capable of matching all observations while employing 
plausible initial conditions.

\subsection{What's coming up?}

Here is a selection (far from complete) of things to look out for in coming 
years which will drive further moves to precision cosmology.

\vspace*{12pt}

\begin{center}
\begin{tabular}{|l|p{8.7cm}|}
\hline
Current & NASA's Map satellite was launched in mid-2001 and is currently making 
an all-sky survey of the CMB (results late 2003??).\\
2002 & Maxima and Boomerang make the first serious attempts to measure CMB 
polarization anisotropies.\\
2001--2004 & Main operations phase of the Sloan Digital Sky Survey \cite{SDSS}, 
seeking to 
redshift a million galaxies.\\
2002--2005 & First systematic surveys for high-redshift galaxy clusters using 
X-rays and the Sunyaev--Zel'dovich effect.\\
2007 & ESA's {\em Planck} satellite launched, for high-resolution all-sky 
mapping of CMB temperature and polarization.\\
2010?? & Launch of the LISA satellites, capable of probing a stochastic 
gravitational wave background (though not the inflationary one except in 
exceptional models).\\
\hline
\end{tabular}
\end{center}

\section{The Inflationary Cosmology}

\subsection{Overview and models}

This section focusses on the last two of the goals listed at the start of this 
article. The claim is that during the very early Universe, a physical process 
known as {\bf inflation} took place, which still manifests itself in our present 
Universe via the perturbations it left behind which later led to the development 
of structure in the Universe. By studying those structures, we hope to shed 
light on whether inflation occurred, and by what physical mechanism. An 
extensive account of inflation appears in my textbook with David Lyth \cite{LL}.

I begin by defining inflation. The scale factor of the Universe at a given time 
is measured by the scale factor $a(t)$. In general a homogeneous and isotropic 
Universe has two characteristic length scales, the curvature scale and the 
Hubble 
length. The Hubble length is more important, and is given by
\begin{equation}
cH^{-1} \quad {\rm where} \quad H \equiv \frac{\dot{a}}{a} \,.
\end{equation}
Typically, the important thing is how the Hubble length is changing with time as 
compared to the expansion of the Universe, i.e.~what is the behaviour of the 
{\bf comoving Hubble length} $H^{-1}/a$?

During any standard evolution of the Universe, such as matter or radiation 
domination, the comoving Hubble length increases. It is then a good estimate of 
the size of the observable Universe. {\bf Inflation} is defined as any epoch of 
the Universe's evolution during which the comoving Hubble length is decreasing
\begin{equation}
\frac{d\left(H^{-1}/a\right)}{dt} < 0 \Longleftrightarrow \ddot{a}>0 \,,
\end{equation}
and so inflation corresponds to any epoch during which the Universe has 
accelerated expansion. During this time, the expansion of the Universe outpaces 
the growth of the Hubble radius, so that physical conditions can become 
correlated on scales much larger than the Hubble radius, as required to solve 
the horizon and flatness problems.

As discussed in the last section, there is very good evidence from observations 
of Type Ia 
supernovae that the Universe is {\em presently} accelerating \cite{Sn}. This is 
usually 
attributed to the presence of a cosmological constant. This is clearly at some 
level good news for those interested in the possibility of inflation in the 
early Universe, as it indicates that inflation is possible in principle, and 
certainly that any purely theoretical arguments which suggest inflation is not 
possible should be treated with some skepticism.

If the Universe contains a fluid, or combination of fluids, with energy density 
$\rho$ and pressure $p$, then
\begin{equation}
\ddot{a}>0 \Longleftrightarrow \rho + 3p < 0 \,,
\end{equation}
(where the speed of light $c$ has been set to one). As we always assume a 
positive energy density, inflation can only take place if the Universe is 
dominated by a material which can have a negative pressure. Such a material is a 
scalar field, usually denoted $\phi$. A homogeneous scalar field has a kinetic 
energy and a potential energy $V(\phi)$, and has an effective energy density and 
pressure given by
\begin{equation}
\rho = \frac{1}{2} \dot{\phi}^2 + V(\phi) \quad ; \quad p = \frac{1}{2} 
\dot{\phi}^2 - V(\phi) \,.
\end{equation}
The condition for inflation can be satisfied if the potential dominates.

A model of inflation typically amounts to choosing a form for the potential, 
perhaps supplemented with a mechanism for bringing inflation to an end, and 
perhaps may involve more than one scalar field. In an ideal world the potential 
would 
be predicted from fundamental particle physics, but unfortunately there are many 
proposals for possible forms. Instead, it has become customary to assume that 
the potential can be freely chosen, and to seek to constrain it with 
observations. A suitable potential needs a flat region where the potential can 
dominate the kinetic energy, and there should be a minimum with zero potential 
energy in which inflation can end. Simple examples include $V = m^2 \phi^2/2$ 
and $V = \lambda \phi^4$, corresponding to a massive field and to a 
self-interacting field respectively. Modern model building can get quite 
complicated --- see Ref.~\cite{LR} for a review.

\subsection{Inflationary cosmology: perturbations}

By far the most important aspect of inflation is that it provides a possible 
explanation for the origin of cosmic structures. The mechanism is fundamentally 
quantum mechanical; although inflation is doing its best to make the Universe 
homogeneous, it cannot defeat the uncertainty principle which ensures that 
residual inhomogeneities are left over. These are 
stretched to astrophysical scales by the inflationary expansion. Further, 
because these are determined by fundamental physics, their magnitude can be 
predicted independently of the initial state of the Universe before inflation. 
However, the magnitude does depend on the model of inflation; different 
potentials predict different cosmic structures.

One way to think of this is that the field experiences a quantum `jitter' as it 
rolls down the potential. The observed temperature fluctuations in the cosmic 
microwave background are one part in $10^5$, which ultimately means that the 
quantum effects should be suppressed compared to the classical evolution by this 
amount.

Inflation models generically predict two independent types of perturbation:
\begin{description}
\item[Density perturbations $\delta_{{\rm H}}^2(k)$:] These are caused by 
perturbations in the scalar field driving inflation, and the corresponding 
perturbations in the space-time metric.
\item[Gravitational waves $A_{{\rm T}}^2(k)$:] These are caused by perturbations 
in the space-time metric alone.
\end{description} 
They are sometimes known as scalar and tensor perturbations respectively, 
because of the way they transform. Density perturbations are responsible for 
structure formation, but gravitational waves can also affect the microwave 
background.

We do not expect to be able to predict the precise locations of cosmic 
structures from first principles (any more than one can predict the precise 
position of a quantum mechanical particle in a box). Rather, we need to focus on 
statistical measures of clustering. Simple models of inflation predict that the 
amplitudes of waves of a given wavenumber $k$ obey gaussian statistics, with the 
amplitude of each wave chosen independently and randomly from a gaussian. What 
it does predict is how the width of the gaussian, known as its amplitude, varies 
with scale; this is known as the {\bf power spectrum}.

With current observations it is a good approximation to take the power spectra 
as being power laws with scale, so
\begin{eqnarray}
\delta_{{\rm H}}^2(k) & = & \delta_{{\rm H}}^2(k_0) \left[ \frac{k}{k_0} 
\right]^{n-1} \\
A_{{\rm T}}^2(k) & = & A_{{\rm T}}^2(k_0) \left[ \frac{k}{k_0} \right]^{n_{{\rm 
T}}}
\end{eqnarray}
In principle this gives four parameters --- two amplitudes and two spectral 
indices --- but in practice the spectral index of the gravitational waves is 
unlikely to be measured with useful accuracy, which is rather disappointing as 
the simplest inflation models predict a so-called consistency relation relating 
$n_{{T}}$ to the amplitudes of the two spectra, which would be a distinctive 
test of inflation. The assumption of power-laws for the spectra requires 
assessment both in extreme areas of parameter space and whenever observations 
significantly improve.

\subsection{The current status of inflation}

The best available constraints come from combining data from different sources; 
for two recent attempts see Wang et al.~\cite{WTZ} and Efstathiou et 
al.~\cite{Eetal}. Suitable data include observations of the recent dynamics of 
the Universe using Type Ia supernovae, cosmic microwave anisotropy data, and 
galaxy correlation function data.

Currently inflation is a definite qualitative success, with striking agreement 
between the predictions of the simplest inflation models and observations. In 
particular, the locations of the microwave anisotropy power spectrum peaks are 
most simply interpreted as being due to an adiabatic initial perturbation 
spectrum in a spatially-flat Universe. The multiple peak structure strongly 
suggests that the perturbations already existed at a time when their 
corresponding scale was well outside the Hubble radius. No unambiguous evidence 
of nongaussianity has been seen.

Quantitatively, however, things have some way to go. At present the best that 
has been done is to try and constrain the parameters of the power-law 
approximation to the inflationary spectra. The gravitational waves have not been 
detected and so their amplitude has only an upper limit and their spectral index 
is not constrained at all. The current situation can be summarized as follows.
\begin{description}
\item[Amplitude $\delta_{{\rm H}}$:] COBE determines this (assuming no 
gravitational waves) to about ten percent accuracy (at one-sigma) as 
approximately
$\delta_{{\rm H}} = 1.9 \times 10^{-5} \, \Omega_0^{-0.8}$ on a scale close to 
the present Hubble 
radius (see Refs.~\cite{BLW,LL} for accurate formulae).
\item[Spectral index $n$:] This is thought to lie in the range $0.8 < n < 1.05$ 
(at 95\% confidence). It would be extremely interesting were the 
scale-invariant case, $n=1$, to be convincingly excluded, as that would be clear 
evidence of dynamical processes at work, rather than symmetries, in creating the 
perturbations.
\item[Gravitational waves $r$:] Measured in terms of the relative contribution 
to large-angle microwave anisotropies, the tensors are currently constrained to 
be no more than about 30\%.
\end{description}

\subsection{Inflation and CMB oscillations}

A key property of inflationary perturbations is that they were created in the 
early 
Universe and evolved freely from then. Although a general solution to the 
perturbation equations has two modes, growing and decaying, only the growing 
mode will remain by the time the perturbation enters the horizon. This leads 
directly to the prediction of an oscillatory structure in the microwave 
anisotropy power spectrum, as seen in Figure~\ref{f:scdm}.\cite{peaks} The 
existence of such 
a structure is a robust prediction of inflation; if it is not seen then 
inflation cannot be the sole origin of structure.

The most significant recent development in observations pertaining to inflation 
is the first clear evidence for multiple peaks in the spectrum, seen by the DASI 
\cite{DASI} and Boomerang \cite{Boom01} experiments shown in 
Figure~\ref{f:datafig}. This is a crucial 
qualitative test which inflation appears to have passed, and which could have 
instead provided evidence against the entire inflationary paradigm for structure 
formation. These 
observations lend great support to inflation, though it must be stressed that 
they are not able to `prove' inflation, as it may be that there are other ways 
to 
produce such an oscillatory structure \cite{Neil}.

\subsection{Prospects for the future}

It remains possible that future observations will slap us in the face and lead 
to inflation being thrown out. But if not, we can expect an incremental 
succession of better and better observations, culminating (in terms of 
currently-funded projects) with the {\em Planck} satellite \cite{Planck}. Faced 
with observational data of exquisite quality, an initial goal will be to test 
whether the simplest models of inflation continue to fit the data, meaning 
models with a single scalar field rolling slowly in a potential $V(\phi)$ which 
is then to be constrained by observations. If this class of models does remain 
viable, we can move on to reconstruction of the inflaton potential from the 
data.

{\em Planck}, currently scheduled for launch in February 2007, should be highly 
accurate. In particular, it should be able to measure the spectral index to an 
accuracy better than $\pm 0.01$, and detect gravitational waves even if they are 
as little as 10\% of the anisotropy signal. In combination with other 
observations, these limits could be expected to tighten significantly further, 
especially the tensor amplitude. Such observations would rule out almost all 
currently known inflationary models. Even so, there will be considerable 
uncertainties, so it is important not to overstate what can be achieved.

\begin{figure}[p]
\centering 
\leavevmode\epsfysize=7.3cm \epsfbox{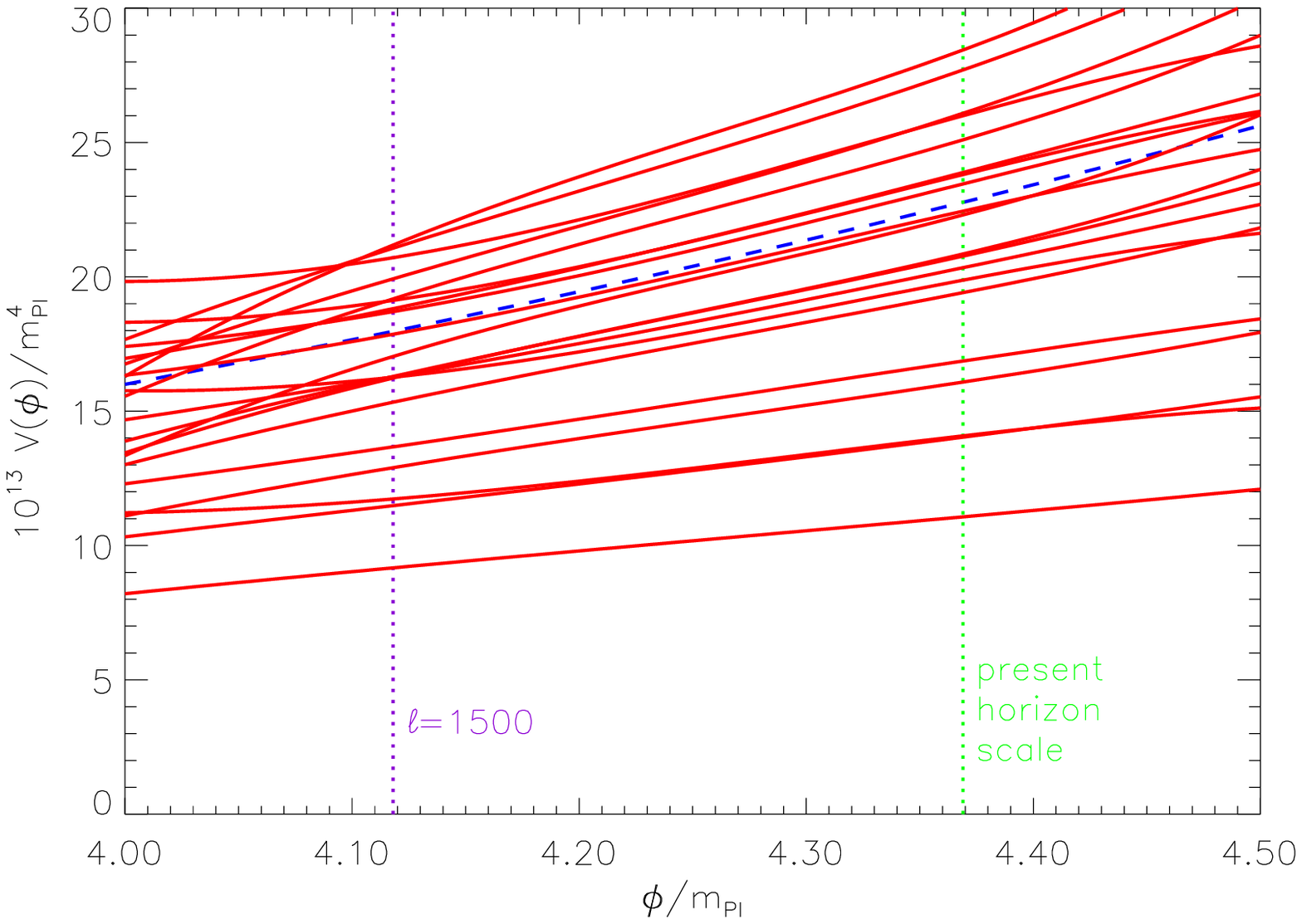}\\ 
\leavevmode\epsfysize=7.3cm \epsfbox{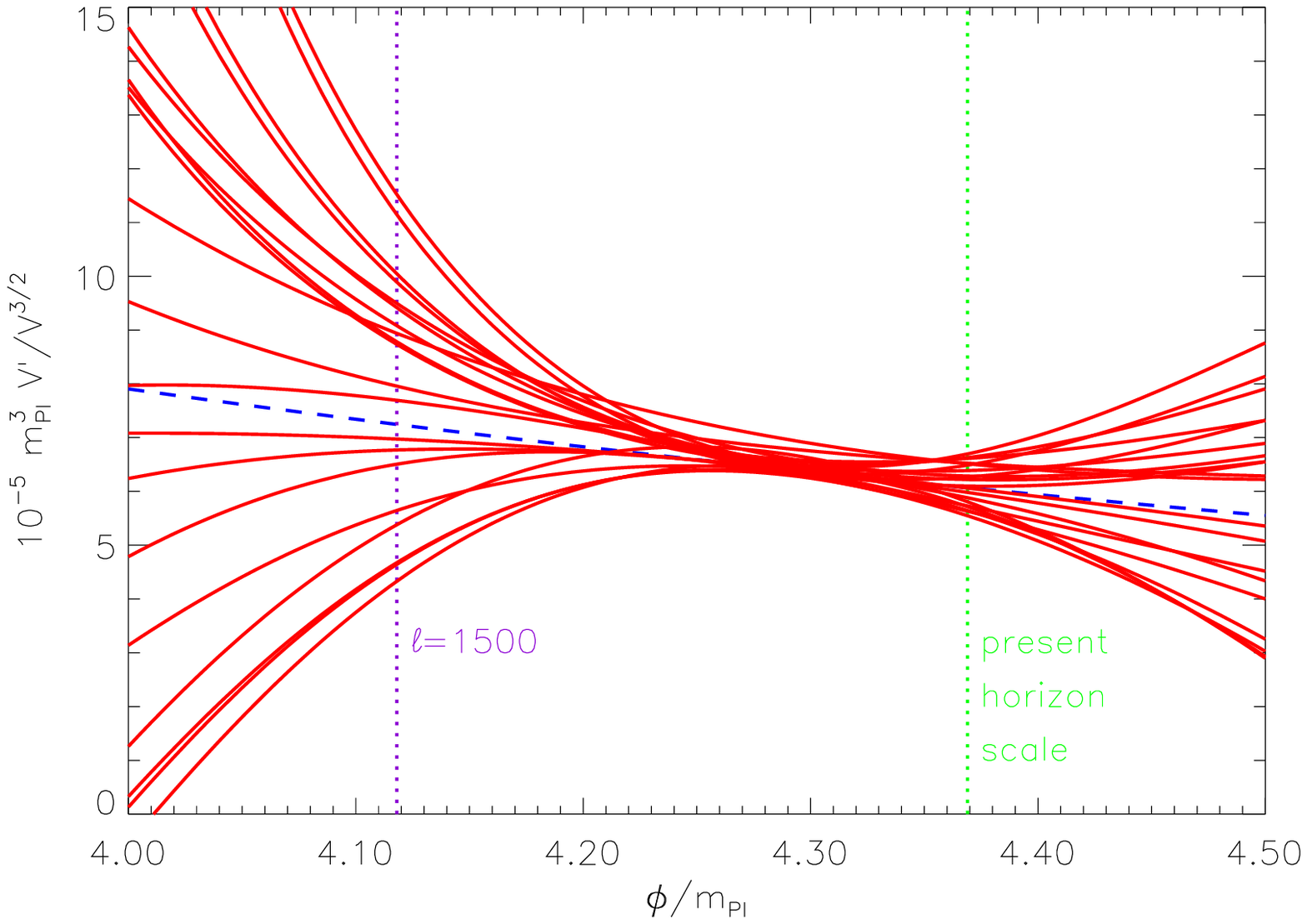}\\ 
\caption[scdm]{\label{f:recon} Sample reconstruction of a potential, where the 
dashed line shows the true potential and the solid lines are thirty Monte Carlo 
reconstructions (real life can only provide one). The upper panel shows the 
potential itself which is poorly determined. However some combinations, such as 
$(dV/d\phi)/V^{3/2}$ shown in the lower panel, can be determined at an accuracy 
of a few percent. See Ref.~\cite{GL} for details.}
\end{figure} 

Reconstruction can only probe the small part of the potential where the field 
rolled while generating perturbations on observable scales. We know enough about 
the configuration of the {\em Planck} satellite to be able to estimate how well 
it should perform. Ian Grivell and I recently described a numerical technique 
which gives an optimal construction \cite{GL}. Results of an example 
reconstruction are shown in Figure~\ref{f:recon}, where it was assumed that the 
true potential was $\lambda \phi^4$. The potential itself is not well determined 
here (the tensors are only marginally detectable), but certain combinations, 
such as $(dV/d\phi)/V^{3/2}$, are 
accurately constrained and would lead to high-precision constraints on inflation 
model parameters.

\section{Selected topics}

The previous sections overviewed the status of two areas of cosmology. In this 
section, I give a short account of some particular topics which have received a 
lot of attention lately.

\subsection{Does cold dark matter really work?}

Of the topics in this article, perhaps the one with the greatest potential 
significance for elementary particle physicists is this one: Is the dark matter 
really cold? Until lately the opinion of the astrophysics community was united 
behind this assumption, based on the remarkable success with which the cold dark 
matter model explains many observations in structure formation. However, more 
recently 
questions have opened up as to whether or not the cold dark matter assumption 
gives a good fit to observations on short scales, and in particular to the 
structure within galaxies.

High-resolution simulations of galaxy formation indicate that the dark matter 
retains considerable clumpiness when the small structures which first form are 
absorbed into larger structures. The persistence of substructure is a success 
for explaining the structure of galaxy clusters, where thousands of galaxies can 
retain their identity upon assimilation, but fails dismally in explaining our 
own galaxy where only a handful of dwarf satellites are observed \cite{subst}. 
Even if they were stripped of their visible baryon components, such knots of 
dark matter would be sufficient to destroy the observed thin disks of spiral 
galaxies.

Potentially related to this are two further problems:
\begin{itemize}
\item Dwarf galaxy cores: theory predicts that the density diverges towards the 
centre of halos, whereas in well-observed dwarf galaxies a uniform-density core 
is seen.
\item Bulge constitution: enough microlensing events have been seen towards the 
Milky Way bulge to suggest that they explain {\em all} the dark matter in the 
central regions of our galaxy, leaving no room for particle dark matter.
\end{itemize}

It remains unclear whether these problems are really so robust that the cold 
dark matter paradigm is under serious threat. However they have been taken 
sufficiently seriously as to motivate a slew of papers on alternative sark 
matter properties, including warm dark matter, self-interacting dark matter, 
annihilating dark matter or condensated dark matter. Whether any of these could 
provide a unified solution to the problems listed above is unclear, but needless 
to say all have major consequences for dark matter search strategies and 
particle physics phenomenology.

\subsection{Do neutrinos play a role?}

What role can neutrinos play in cosmology? This is a topical question as 
evidence mounts up in favour of a non-zero neutrino mass from solar and 
atmospheric neutrino experiments. 

Assuming standard interactions and negligible lepton asymmetry, theory predicts
\begin{equation}
\Omega_\nu = \frac{\sum_i m_{\nu_i}}{90 \, h^2 \, {\rm eV}}
\end{equation}
Neutrinos could have a measurable effect on structure formation, through their 
free-streaming, even if $\Omega_\nu$ were as little as 0.01, meaning $m_\nu \sim 
0.5$ eV suggesting that neutrinos play only a very minor role in structure 
formation. However recent observations favour even smaller values. However if 
for 
some reason there were a substantial lepton asymmetry, there could be an effect 
at even smaller masses.

It's also worth noting that, even if massless, the behaviour of relic neutrinos 
does have to be taken into account to compute quantities such as the microwave 
anisotropy power spectrum. Observations of it therefore do offer an indirect 
confirmation that the relic neutrino population predicted by theory does in fact 
exist.

\subsection{Are existing treatments of inflation oversimplistic?}

Much of the information disseminated from the inflationary community to the 
broader physics and astronomy community is based around the simplest paradigm, 
where a single scalar field slow-rolls down a potential. While this continues to 
be in excellent agreement with observation, and is a powerful working hypothesis 
worthy of testing in its own right, much work has recently gone into studying 
more complicated situations. It is a continuing challenge to uncover the full 
phenomenology of models with more than one dynamical field. Such a situation 
changes many of the usual assumptions. There is no longer a unique trajectory 
for the inflation field, and predictions for the density perturbations may well 
become dependent on initial conditions. Perhaps more importantly, the 
perturbations are no longer guaranteed to be adiabatic, and isocurvature 
perturbations may well be non-gaussian and/or correlated with the adiabatic 
component. If the single-field paradigm fails, it will be important to 
understand whether there remain well-motivated inflationary models capable of 
explaining the data, particularly as efforts to determine cosmological 
parameters such as $h$ and $\Omega_{{\rm B}}$ will flounder if the initial 
perturbations cannot be accurately parametrized.

\subsection{The braneworld}

At a particle physics conference or school you can quite happily state that `it 
is generally accepted that our Universe has more than three spatial dimensions' 
without much fear of contradiction, though it is probably not necessary even to 
step out of a physics building to find out how limited this general acceptance 
actually is. Nevertheless, extra dimensions have been with particle theorists 
consistently for a long time now, and undoubtedly they are an issue which may be 
of considerable importance for early Universe cosmology.

Until recently, it had been assumed that the failure to observe the predicted 
extra dimensions meant that the extra ones were ``curled up'' to be unobservably 
small. However, there is a now a new idea, the {\bf 
braneworld}, which proposes that at least one of these extra dimensions might be 
relatively large, with us constrained to live on a three-dimensional {\bf brane} 
running 
through the higher-dimensional space. Gravity is able to propagate in the full 
higher-dimensional space, which is known as the {\bf bulk}.

This radical idea has many implications for cosmology, both in the present and 
early Universe, and so far we have only scratched the surface of possible new 
phenomena. Already many exciting results have been obtained; here there is only 
space to consider a few pertinent questions.

\vspace*{6pt}
\noindent
{\em a)~Are there modifications to the evolution of the homogeneous Universe?}\\
The answer appears to be yes; for example in a simple scenario (known as 
Randall--Sundrum Type II \cite{RSII}) the Friedmann equation is modified at high 
energies so that, after some simplifying assumptions, it reads \cite{bin}
\begin{equation}
H^2 = \frac{8\pi G}{3} \left( \rho+ \frac{\rho^2}{2\lambda} \right) \,,
\end{equation} 
where $\lambda$ is the tension of the brane. This recovers the usual cosmology 
at low energies $\rho \ll \lambda$, but otherwise we have new behaviour. This 
opens new opportunities for model building, see for example Ref.~\cite{CLL}.

\vspace*{6pt}
\noindent
{\em b)~Are inflationary perturbations different?}\\
Again the answer is yes --- there are modifications to the formulae giving 
scalar and tensor perturbations \cite{braneperts}. Unfortunately the main effect 
of this is to introduce new degeneracies in interpreting observations, as a 
potential can always be found matching observations for any value of $\lambda$ 
\cite{LT}. The initial perturbations therefore cannot be used to test the 
braneworld scenario.

\vspace*{6pt}
\noindent
{\em c)~Do perturbations evolve differently after they are laid down on large 
scales?}\\
The answer here is less clear. It is possible that perturbation 
evolution is modified even at late times, e.g.~perturbations in the bulk 
could influence the brane in a way that couldn't be predicted from brane 
variables alone. Whether there is a significant effect is unclear and is likely 
to be model dependent.

\subsection{The Ekpyrotic Universe}

It has recently been proposed that the Big Bang is actually the result of the
collision of two branes, dubbed the Ekpyrotic Universe \cite{ekpyrotic}.  It has
been claimed that this scenario can provide a resolution to the horizon and
flatness problems, essentially because causality arises from the
higher-dimensional theory and allows a simultaneous Big Bang everywhere on our
brane, though existing implementations solve the problem by hand in the initial
conditions.  As I write this, it remains unclear how to successfully describe
the instant of collision between the two branes (the singularity problem), and
considerable controversy surrounds whether or not the scenario can also generate
nearly scale-invariant adiabatic perturbations \cite{ekperts}.  Both aspects are
required to make it a serious rival to inflation.\footnote{Since the talks on 
which this article is based were given, there is also a newer incarnation of the 
scenario known as the Cyclic Universe \cite{cyclic} which further extends these 
ideas.}

\section*{Acknowledgments}
The author was supported in part by the Leverhulme Trust.

\end{document}